\documentclass[aps,nofootinbib,preprintnumbers,citeautoscript]{revtex4}
\usepackage{amsmath,amssymb}
\usepackage{enumerate}
\usepackage{graphicx}
\graphicspath{{/home/kapil/My_data/kapilpaper/arXiv/DM_Int_as_an_agent/figures/}}

\date{\today}

\begin{document}
\title{Dzyaloshinskii-Moriya interaction as an agent to free the bound entangled states}

\author{Kapil K. Sharma$^\ast$ and  S. N. Pandey$^\dagger$  \\
\textit{Department of Physics, Motilal Nehru National Institute of Technology,\\ Allahabad 211004, India.} \\
E-mail: $^\ast$iitbkapil@gmail.com, $^\dagger$snp@mnnit.ac.in}

\begin{abstract}
In the present paper, we investigate the efficacy of Dzyaloshisnhkii-Moriya (DM) interaction to convert the bound entangled states into free entangled states. We consider the tripartite hybrid system as a pair of non interacting two qutrits initially prepared in bound entangled states and one auxiliary qubit. Here we consider two types of bound entangled states investigated by Horodecki. The auxiliary qubit interacts with any one of the qutrit of the pair through DM interaction. We show that by tuning the probability amplitude of auxiliary qubit and DM interaction strength one can free the bound entangled states, which can be further distilled. We use the reduction criterion to find the range of the parameters of probability amplitude of auxiliary qubit and DM interaction strength, for which the states are distillable. The realignment criterion and negativity have been used for detection and quantification of entanglement.
\end{abstract}
\maketitle



\section{Introduction}
Quantum entanglement \cite{EPR1935,Neilsen2000} is a physical phenomenon which takes place when particles interact in microscopic world in such a way that the quantum state of one particle can be described in terms of each other. The entanglement phenomenon is expected to be a useful resource for future quantum technologies. Many applications, based on entanglement, have been investigated, like quantum teleportation \cite{CBennet1933}, quantum imaging \cite{qi}, quantum game theory \cite{qs}, secure key quantum transmission \cite{AkEkert1991}, etc. To develop quantum technologies, based on entanglement, the quantum community needs long time entangled
quantum systems, which are free from noise. Quantum systems are too evasive as they may loose their entanglement by external environmental interactions and can go for entanglement sudden death (ESD) for finite time \cite{YuEberly2004,YuEberly2009}. So, dynamics of entanglement and its control under various environmental interactions conceptually underpinning the quantum information processing. Dynamics in various quantum spin chains have been studied under Dzyaloshisnhkii-Moriya (DM) interaction \cite{DMmoriya,Tmoriya2_1960}. DM interaction is a useful resource in quantum information processing to entangle and disentangle the quantum systems. Recently, Zang \textit{et al.} studied the entanglement dynamics of two qubit pair by taking an third qubit or qutrit, which interact with a qubit of the pair through DM interaction \cite{ZQiang1,ZQiang2,ZQiang3}. They studied the dynamics of entanglement by taking a third qubit as a controller qubit. Further, they studied the same by taking a third qutrit as a controller qutrit. They proposed that by manipulating the probability amplitude of third qubit or qutrit and DM interaction strength one can control the entanglement between two qubits induced by DM interaction and hence in various quantum spin chains \cite{Hisenberg}. At this point, we mention here that the same method may not only be used to control the entanglement, but it may also be used to free the bound entanglement \cite{Horodecki_11,Horodecki_12,Horodecki_21,Horodecki_22} in various bound entangled states. Once the bound entangled states are free, then they can be distilled \cite{Dist}.

The quantum states beyond the dimension $3\otimes3$ have been classified in two categories like free entangled states and bound entangled states. Free entangled states are distillable states or in other words noise free states. On the other hand, bound entangled states are noisy states and no pure entanglement can be obtained by local operations and classical communication. It is difficult to use  bound entangled states directly for practical quantum information processing. However, by providing additional resource, bound entanglement can be activated to increase the fidelity of quantum teleportation \cite{Activation_1,Activation_2,Activation_3}. Up to now we don't have satisfactory tools to quantify and detecting  the bound entanglement. Recently, the free entanglement production from bound entangled states is proposed by using the ancillary system which is coupled to the initial bound-entangled state via appropriate weak measurements \cite{Activation_4}.

In the present work, we consider qutrit-qutrit bound entangled bipartite states proposed by Horodecki \cite{Horodecki_11,Horodecki_12,Horodecki_21,Horodecki_22}. The main goal of the present work is to show that DM interaction can be a used as a useful agent to free the bound entangled state in two qutrit system. Once the states are free, further these can be distilled. We have used reduction criterion to check the distillability of these states. Using this criterion, one can ensure that the states are surely distillable. Further, we have used realignment criterion to detect the bound entanglement and negativity to measure the free entanglement. The motivation of this study comes from our recent works  \cite{KK1,KK2,KK3,KK4}. 

The plan of the paper is as follows. In Sect. 2, we present the Hamiltonian of the system. Sect 3. is devoted to the description of Horodecki’s bound entangled states and reduced system dynamics. In Sect. 4, we discuss the reduction criterion, realignment criterion and negativity. Sect 5. is devoted to time evolution of negativity and realignment criterion. Lastly in Sect. 6, we report our conclusion.
\begin{figure*}
        \centering
                \includegraphics[width=0.6\textwidth]{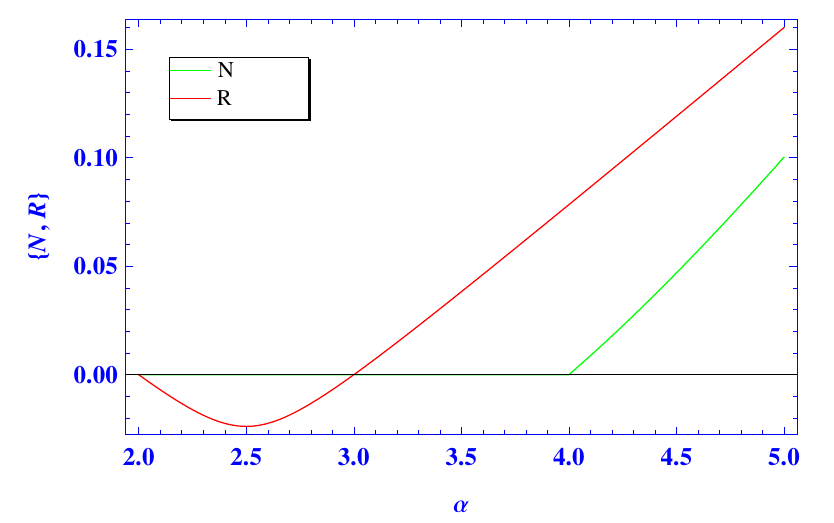}
                \caption{Plot of entanglement and realignment criterion with $D = 0.0$. Green color graph represents negativity (N) and red color graph represents the realignement criterion (R).} \label{1}
\end{figure*}
\section{Hamiltonian of the system}
We consider a qutrit (A)-qutrit (B) pair and an auxiliary qubit (C) which interact with any one of the qutrit of the pair through DM interaction. Here, we assume that the auxiliary qubit (C) interact with the qutrit (B) of the pair. Now the Hamiltonian of the system can be written as
\begin{equation}
H=H_{AB}+H_{BC}^{int}, \label{eq:17}
\end{equation}
where $H_{AB}$ is the Hamiltonian of qutrit (A) and qutrit (B) and $H_{BC}^{int}$ is the interaction Hamiltonian of qutrit (B) and qubit (C). We consider uncoupled qutrit (A) and qutrit (B), so $H_{AB}$ is zero. Now the Hamiltonian becomes
\begin{eqnarray}
H=H_{BC}^{int}=\vec{D}.(\vec{\sigma_B} \times \vec{\sigma_C}),   \label{eq:18}
\end{eqnarray}
where $\vec{D}$ is DM interaction between qutrit (B) and qubit (C) and
$\vec{\sigma_B}$, $\vec{\sigma_C}$ are associated vectors of qutrit (B) and qubit (C) respectively. 
We assume that DM interaction exist along the z-direction only. In this case, the Hamiltonian can be simplified as\\ 
\begin{equation}
H=D.(\sigma_B^X \otimes \sigma_C^Y-\sigma_B^Y \otimes \sigma_C^X),    \label{eq:19} \\ 
\end{equation}
where $\sigma_B^X$ and $\sigma_B^Y$ 
are Gell-Mann matrices for qutrit (B) and $\sigma_C^X$ and $\sigma_C^Y$ are X and Y Pauli matrices of qubit (C) respectively. The above Hamiltonian is a matrix having $6\times 6$ dimension. Further, it is multiplied by the identity matrix of dimension three and we obtained the dimension as $18\times 18$. The matrix of Hamiltonian is easy to diagonalize by using the method of eigendecomposition. The unitary time evolution operator is easily commutable  as 
\begin{eqnarray}
U(t)=e^{-i H t}, \label{eq:20}
\end{eqnarray}
which is also a $18 \times 18$ matrix. This matrix has been used to obtain the time evolution of density matrix of the system.
\begin{figure*}
        \centering
                \includegraphics[width=0.6\textwidth]{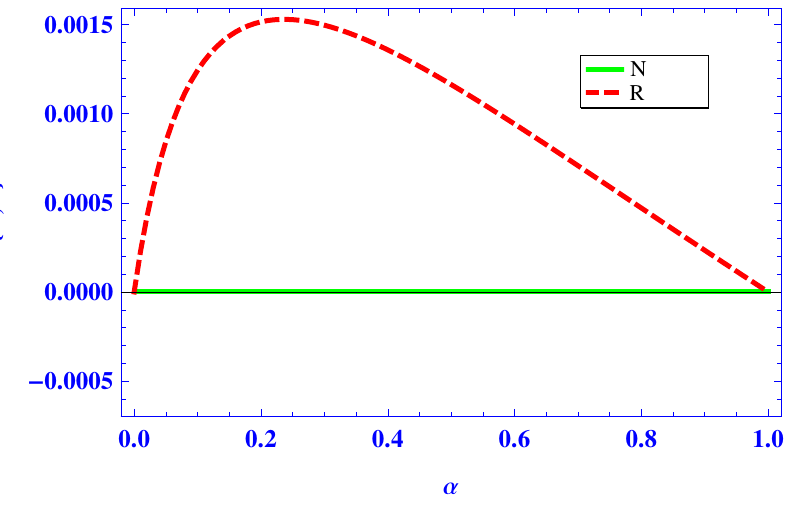}
                \caption{Plot of negativity $(N)$ and realignment criterion $(R)$ with the parameter values $c_{0}=0$ and $Dt=0$. Green color graph represents negativity (N) and red graph represents realignment criterion (R).} \label{4}
\end{figure*}
\section{Horodecki's bound entangled states and reduced system dynamics}
In this section, we describe two known bipartite Horodecki's bound entangled states \cite{Horodecki_11,Horodecki_12,Horodecki_21,Horodecki_22} in $3 \otimes 3$ dimension. In subsection 3.1, we present the state 1 and in subsection 3.2 the state 2 is presented.
\subsection{State 1.} \label{s1}
The Horodecki's bound entangled state \cite{Horodecki_11,Horodecki_12} in $3\otimes 3$ dimension is given by 
\begin{eqnarray}
\rho_{\alpha}(0)=\frac{2}{7}P+\frac{\alpha}{21}Q+\frac{(5-\alpha)}{21}R, \quad 2\leq \alpha \leq 5,  \label{eq:b1}
\end{eqnarray}
where
\begin{align}
& P=|\psi\rangle\langle\psi|, \quad |\psi\rangle=\frac{1}{\sqrt{3}}(|00\rangle+|11\rangle+|22\rangle) \\
& Q=(|01\rangle\langle 01|+|12\rangle \langle 12|+|20\rangle\langle 20|), \\
& R=(|10\rangle\langle 10|+|21\rangle \langle 21|+|02\rangle\langle 02|). \label{eq:con}
\end{align}
Horodecki demonstrated that \\
\begin{center}
\begin{equation}
\rho_{\alpha}(0)=
\begin{cases}
\text{Separable for} \quad 2\leq \alpha \leq 3, \\
\text{Bound entangled for} \quad 3< \alpha \leq 4, \\
\text{Free entangled for} \quad 4< \alpha \leq 5. 
\end{cases} \label{eq:condition}
\end{equation}
\end{center} 
\subsection{State 2.} \label{case2}
Another well known bound entangled state investigated by Horodecki \cite{Horodecki_21,Horodecki_22} in $3 \otimes 3$ dimension is given as\\
\begin{center}
\begin{equation}
\varrho_\alpha={1 \over 8\alpha + 1}
\left[ \begin{array}{ccccccccc}
          \alpha &0&0&0&\alpha &0&0&0& \alpha   \\
           0&\alpha& 0&0&0&0&0&0&0     \\
           0&0&\alpha&0&0&0&0&0&0     \\
           0&0&0&\alpha&0&0&0&0&0     \\
          \alpha &0&0&0&\alpha&0&0&0& \alpha     \\
           0&0&0&0&0&\alpha&0&0&0     \\
           0&0&0&0&0&0&{1+\alpha \over 2}&0&{\sqrt{1-\alpha^2} \over 2}\\
           0&0&0&0&0&0&0&\alpha&0     \\
          \alpha &0&0&0&a&0&{\sqrt{1-\alpha^2} \over 2}&0&{1+\alpha \over 2}\\
       \end{array}
      \right ], \ \ \
\end{equation} \label{b22}
\end{center}
here $0<\alpha<1$.
\subsection{Reduced system dynamics}
In this subsection, we obtain reduced density matrix of qutrit-qutrit system. To begin with, let us consider the auxiliary qubit (C) in pure state as
\begin{equation}
|\phi\rangle=c_0|0\rangle+c_1|1\rangle \label{eq:21}
\end{equation}
with normalization condition 
\begin{equation}
|c_0|^2+|c_1|^2=1 \label{eq:normal}   
\end{equation}
where the probability amplitudes $c_{0}$ and $c_{1}$ are complex in general. \\
Qutrit (A)-qutrit (B) is prepared initially in bound entangled state. The density matrix of bound entangled state before interaction with qubit (C) is given by  (\ref{eq:b1}). Now we can obtain the composite density matrix of the open system, $\rho_{comp.}(0)$ as 
\begin{equation}
\rho_{comp.}(0)=\rho_{\alpha}(0)\otimes \rho_c,  \label{eq:25}
\end{equation}
where $\rho_c$ is the density matrix of the auxiliary qubit (C). The density matrix after the interaction at time $t$ is given by 
\begin{equation}
\rho_{comp.}(t)=U(t)\rho_{comp.}(0)U^\dagger(t), \label{eq:26}
\end{equation}
where U(t) is unitary time evolution operator given by (\ref{eq:20}). Now we make the order same of all the matrices involved in equation (14) as $18\times 18$. This can be done by doing the tensor product of $\rho_{comp.}(0)$ with identity matrix of the order $2\times 2$. Now the order of the matrix  $\rho_{com.}(t)$ becomes as $18\times 18$. Taking partial trace of $\rho_{com.}(t)$ over the basis of auxiliary qubit (C), we get the reduced density matrix $\rho^{AB}$ of $9\times 9$ dimension as,
\begin{equation}
\rho^{AB}=Ptr_c[\rho_{comp.}(t)]. \label{eq:reduce_density}
\end{equation}
Now we obtain the reduced density matrix for bound entangled state proposed by the Horodecki given by Eq. (\ref{eq:b1}). The reduced density matrix is given below 
\begin{center}
\begin{equation}
\rho^{AB}=\left[\begin{array}{ccccccccc}
X_{11}	& 0		    & X_{13}	& 0     	& X_{15}	& 0  	 	 & X_{17}	& 0 		& X_{19}\\
0	    & X_{22}	& 0			& X_{24} 	& 0			& 0      	 & 0 		& 0 		& 0     \\  
X_{31}	& 0			& X_{33}	& 0			& X_{35}	& 0	     	 & 0		& 0 		& X_{39} \\
0		& X_{42}	& 0			& X_{44}	& 0			& 0	     	 & 0 		& 0 		& 0      \\
X_{51}	& 0			& X_{53}	& 0			& X_{55} 	& 0	 		 & X_{57}	& 0  		& X_{59} \\
0		& 0			& 0			& 0			& 0			& X_{66}	 & 0 		& X_{68} 	& 0      \\ 
X_{71}	& 0			& 0			& 0			& X_{75}	& 0	 	 	 & X_{77}	& 0 		& X_{79} \\
0		& 0			& 0			& 0			& 0			& X_{86}  	 & 0		& X_{88}	& 0     \\ 
X_{91}	& 0			& X_{93}	& 0			& X_{95}	& 0		 	 & X_{97} 	& 0 		& X_{99} \end{array}
      \right ], \nonumber  
\end{equation} 
\end{center}
where 
\begin{align}
&X_{11}=X_{19}=X_{91}=X_{99}=\frac{2}{21}, &&X_{13}=-X_{17}=X_{31}=-X_{39}=X_{71}=X_{79} \nonumber \\ && &=X_{93}=-X_{97}=\frac{2}{21}p,  \nonumber \\
&X_{15}=X_{51}=X_{59}=X_{95}=\frac{2}{21}q, &&X_{22}=\frac{1}{21}[\alpha c_{0}^2-(\alpha -5)\frac{p^2}{c_{1}^{2}}+\alpha c_{1}^{2} q^2] , \nonumber \\
& X_{24}=X_{42}=-\frac{1}{21}[\alpha+(-5+\alpha)q]p, &&X_{33}=\frac{(7-\alpha)}{42}-(\alpha -3)\frac{s}{c_{0} c_{1}}-2(\alpha -5)c_{1}^{2}, \nonumber \\
&X_{35}=X_{53}=(-3+\alpha)r,  &&X_{44}=-\frac{1}{21}(\alpha -5)c_{0}^{2}+\frac{(5-2\alpha)s c_{1}}{c_{0}}+\frac{5}{42}\nonumber \\
&X_{55}=\frac{1}{42}[(\alpha+2)c_{0}^{2}-(\alpha-7)c_{1}^{2}]-s, && X_{57}=X_{75}=(-2+\alpha)r, \nonumber \\
&X_{66}=\frac{5 c_{0}^{2}}{42}+\frac{(2\alpha-5)s c_{0}}{c_{1}}+\frac{1}{21}\alpha c_{1}^{2}, &&X_{68}=X_{86}=-\frac{1}{21}[-5+\alpha+\alpha q]p, \nonumber \\
&X_{77}=\frac{1}{21}\alpha c_{0}^{2}+[\frac{(\alpha+2)}{42} c_{1}^{2}+\frac{(\alpha-2)s c_{1}}{c_{0}}] , &&X_{88}=\frac{1}{21}[5c_{1}^2-((-5+\alpha)c_{0}^2+\alpha c_{1}^2)q^2] \nonumber \\  \label{reduced}
\end{align} 
and \\
$p=\sin[\sqrt{2}Dt]c_{0}c_{1}$, $q=\cos[\sqrt{2}Dt]$, $r=\frac{1}{42}\sin[2\sqrt{2} Dt]c_{0}c_{1}$, $s=\frac{1}{42}\cos[2\sqrt{2} Dt]c_{0}c_{1}$. \\
Observing the reduced density matrix, we conclude that $c_{0}$, $D$ and $t$ are the key parameters which influence the entanglement between two qutrit pair. Similarly, we obtain the reduced density matrix corresponding to another bound entangled state given in subsection \ref{b22} with the help of Eq. (\ref{eq:reduce_density}).
\begin{figure*}
        \centering
                \includegraphics[width=0.6\textwidth]{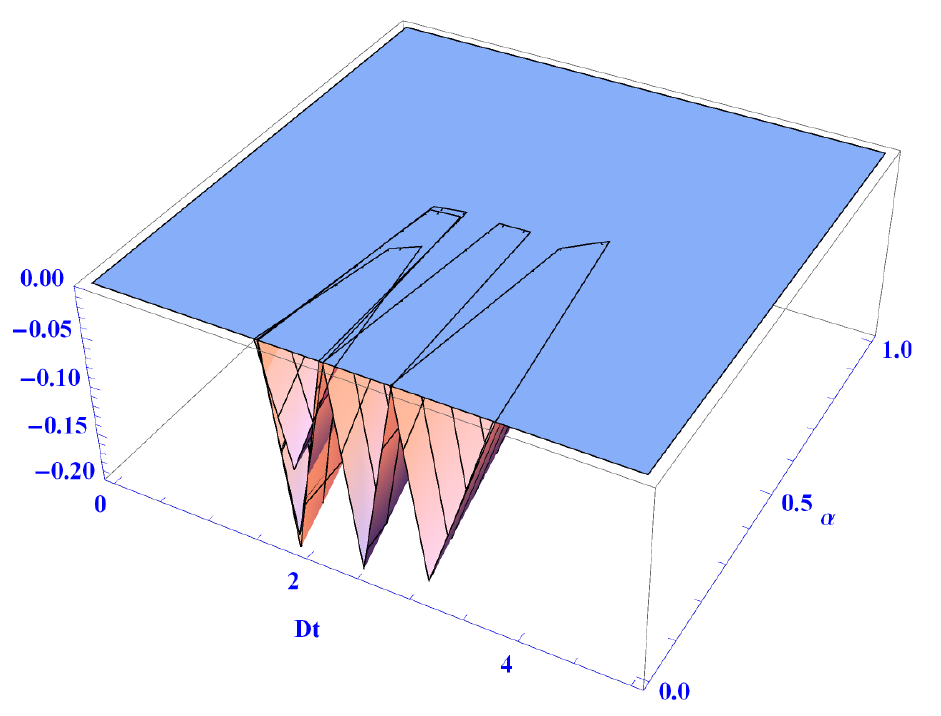}
                \caption{Plot of eigenvalues (Eig.) of reduction criterion with the parameters range $0\leq Dt \leq 5$ and $0 \leq \alpha \leq 1$.}\label{2}
\end{figure*}
\begin{figure*}
        \centering
                \includegraphics[width=0.6\textwidth]{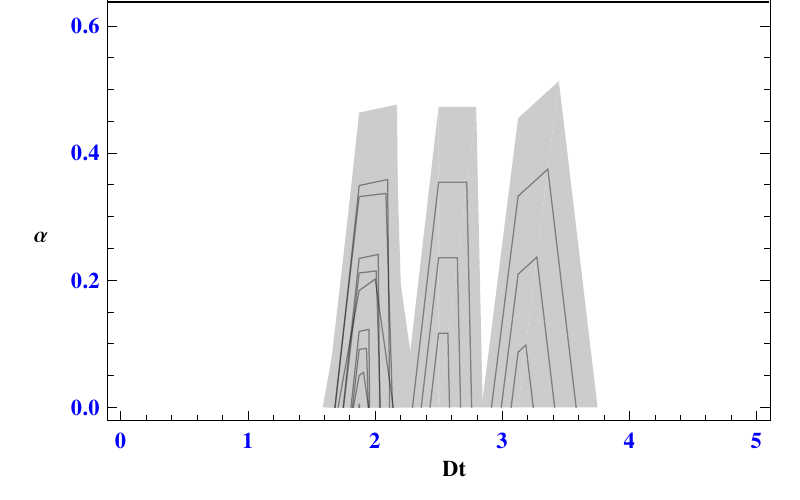}
                \caption{Contour plot of Fig 2, the state is distillable with the parameters range $1.59 \leq Dt \leq$ and $0 \leq \alpha \leq 3.439$}\label{3}
\end{figure*}
\section{Reduction, realignment criterion and negativity}
In this section, we discuss the reduction criterion and its ability to detect the separability and distillability of bipartite states. For the separability of a composite system AB, the following conditions should be satisfied 
\begin{eqnarray}
\rho_{A}\otimes I-\rho \geq 0 \quad \text{and} \quad
I \otimes \rho_{B}-\rho \geq 0, \label{r1}
\end{eqnarray}
where $\rho_{A}=Tr_{B}(\rho)$, $\rho_{B}=Tr_{A}(\rho)$ and $\rho$ is the composite density matrix of the system AB. The states which violates the condition (\ref{r1}) can be distilled. So, for distillability the following condition should be necessarily satisfied \cite{Red}.
\begin{eqnarray}
\rho_{A}\otimes I-\rho \leq 0 \quad \text{or} \quad I\otimes \rho_{B}-\rho \leq 0. \label{r2}
\end{eqnarray} 
Further, we obtain the partial transpose and realigned matrix of the reduced density matrices to obtain the negativity and realignment by using the following matrices
\begin{eqnarray}
(\rho_{ij,kl}^{T})=\rho_{il,kj} \label{t}\\
(\rho_{ij,kl}^{R})=\rho_{ik,jl}. \label{r}
\end{eqnarray}
These matrices have been used to calculate the quantities defined as
\begin{align}
&N=\frac{||\rho^{T}||-1}{2}, &&R=\frac{||\rho^{R}||-1}{2}, \label{formulae}
\end{align} 
where $||.. ||$ is the trace norm of the matrix. The first quantity corresponds to negativity \cite{locc,negativity,Perse1996,A.Peres,schidmt} and it has been used to measure the free entanglement while the second one has been used to detect the bound entanglement in the qutrit-qutrit system. Either $N>0$ or $R>0$ implies that the state is entangled. $N=0$ and $R>0$ implies that the state is PPT bound entangled, and $N>0$ corresponds to free entangled state. However, there is no evidence for the existence of NPT bound entangled states \cite{NPT} as yet, but we can not avoid this future possibility. 
\section{Time evolution of negativity and realignment criterion}
\label{dist}
In this section, first we verify the calculation of reduced density matrices for $(Dt=0)$ for state 1. The reduced density matrix of two qutrits corresponding to state 1 is given by Eq. (\ref{reduced}).
We put the values of the parameters $Dt$ and $c_{0}$ as zero in reduced density matrix then reduced density matrix maps to the initial state of two qutrits given by Eq. (\ref{eq:b1}), which verify the correctness of reduced density matrix obtained in calculation. Further, we plot the negativity and realignment criterion and the result is shown in Fig. \ref{1}. The green color graph represent the negativity (N), while the red color graph represents the realignment criteria (R) in the figure. The realignment criterion for $2\leq \alpha \leq 3$ is negative, so it shows that the state is separable. For $3< \alpha \leq 4$ the realignment criterion is positive, which show that the state is bound entangled. Further, for $4< \alpha \leq 5$ the negativity graph is positive, which shows that the state is free entangled. These results match with the result given by Eq. (\ref{eq:condition}). 

Similarly, we verify the reduced density matrix obtained for the state 2 by putting the parameter values $Dt=0$ and $c_{0}=0$. The reduced density matrix maps to the initial density matrix of the state 2 given in subsection \ref{b22}, which proves that the reduced density matrix obtained is correct. Now, we plot the negativity and realignment criterion both in the absence of DM interaction in Fig. \ref{4}. The realignment criterion achieves positive values for $0\leq \alpha \leq 1$ and negativity is zero. These results show that initially the state 2 of two qutrits is bound entangled with $0\leq \alpha \leq 1$. 
\begin{figure*}
        \centering
                \includegraphics[width=1.0\textwidth]{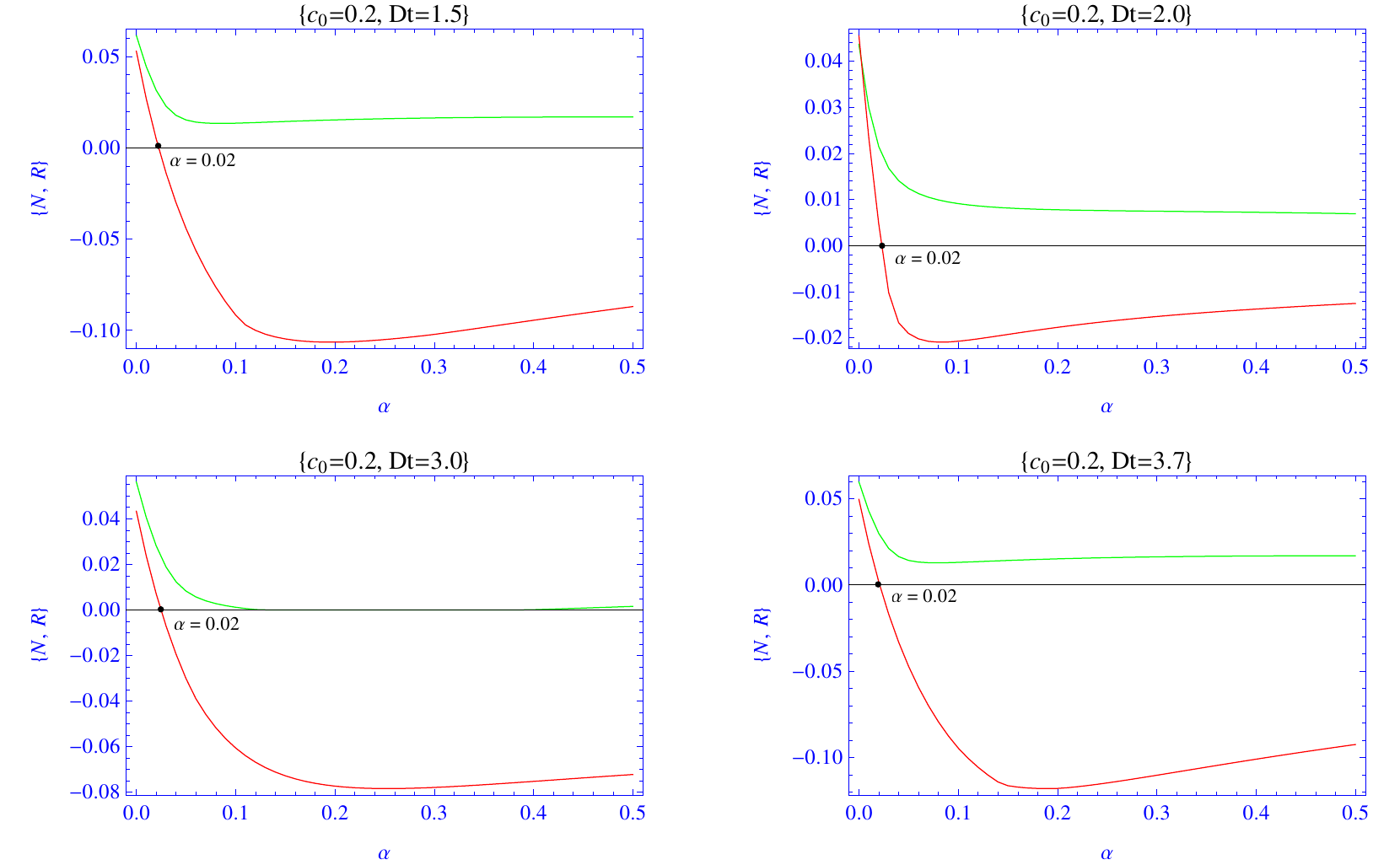}
                \caption{Plot of negativity (N) and realignment criterion (R) with the parameter range $c_{0}=0.2$ and ($1.5 \leq Dt \leq 3.7$, $0\leq \alpha \leq 0.5$). Green color graphs represent N and red graphs represent R. Realignment criterion fails at $\alpha=0.02$.} \label{5}
\end{figure*}
Further, in this section we detect the distillability of the Horodecki bound entangled states 1 and 2, by using the reduction criterion with the reduced density matrices of two qutrits in the presence of DM interaction. Next, we study the evolution of free entanglement and realignment criterion obtained by using Eq. (\ref{formulae}) for the bound entangled states 1 and 2 in these two cases. In case 1, we consider the bound entangled state described under the subsection 3.1 and in case 2, we do the same for another bound entangled state given in subsection 3.2. 
\subsection{Case 1} 
In this case, we consider the state 1 given in subsection 3.1. Here, we replace $c_{1}^{2}$ in Eq. (\ref{formulae}) in terms of $c_{0}^{2}$ by using the normalization condition given in Eq. (\ref{eq:normal}). Next we check the distillability of the state by using the reduction criterion give in Eq. (\ref{r2}). We found that the matrices in reduction criterion, i.e., $(\rho_{A}\otimes I-\rho)$ or $(I\otimes \rho_{B}-\rho)$, incorporate the parameters $Dt$, $c_{0}$ and $\alpha$. Further, the eigenvalues of these matrices have been calculated by taking the range of the parameters as ($3\leq \alpha \leq 4$, $0\leq Dt \leq 5$) with the varying value of the parameter $c_{0}$. With the range ($0\leq c_{0}\leq 1$). It is found that the eigenvalues of these matrices achieve positive values. These eigenvalues are obtained with the maximum values of $c_{0}$ (i.e. $c_{0}=1$), it is found that all eigenvalues are positive, so the reduction criterion given in Eq. (\ref{r2}) fails and state 1 is not distillable. 
\begin{figure*}
        \centering
                \includegraphics[width=1.0\textwidth]{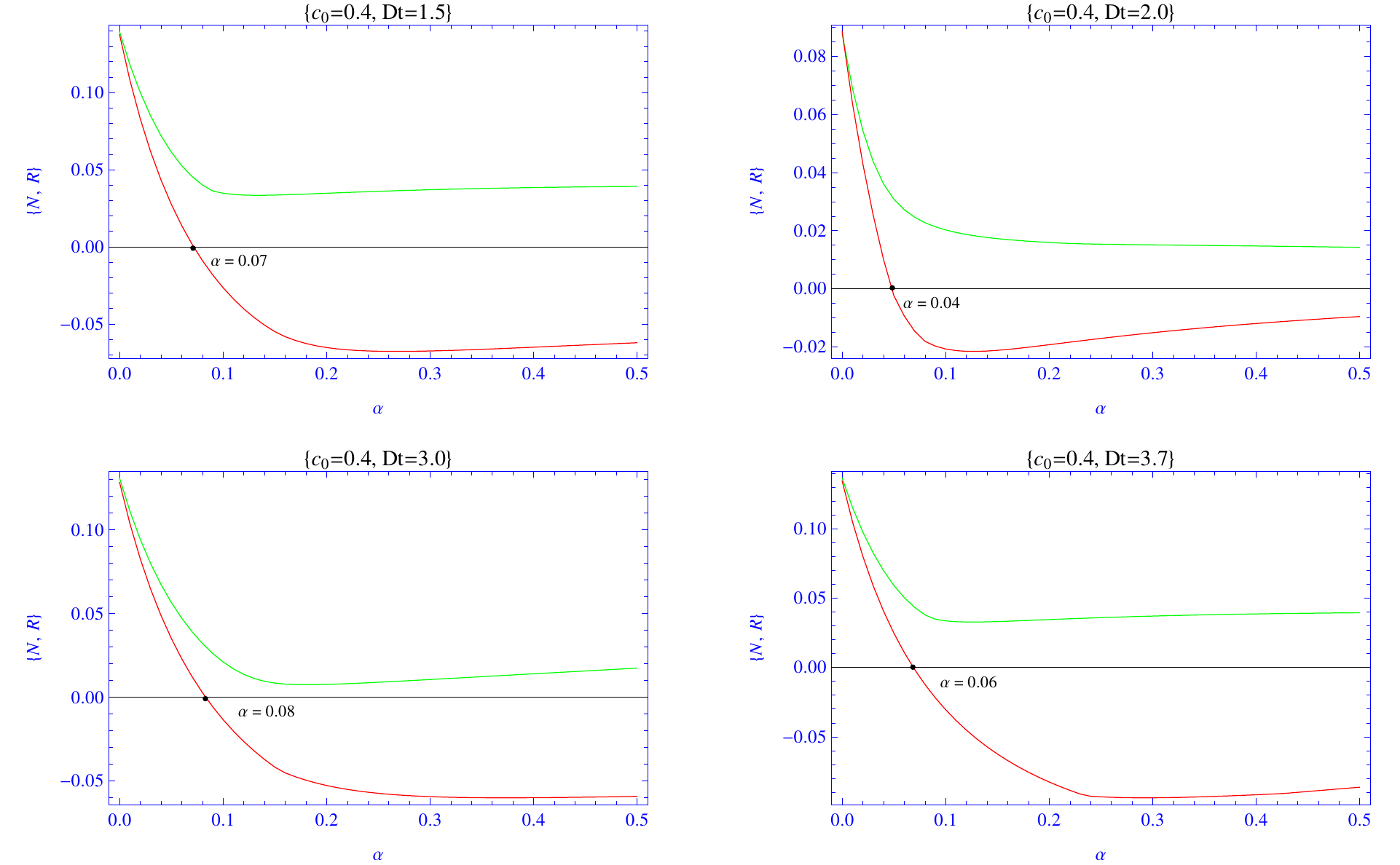}
                \caption{Plot of negativity (N) and realignment criterion (R) with the parameter range $c_{0}=0.4$ and $1.5 \leq Dt \leq 3.7$. Green color graphs represents N and red color graph represents R.}\label{6}
\end{figure*}
\begin{figure*}
        \centering
                \includegraphics[width=1.0\textwidth]{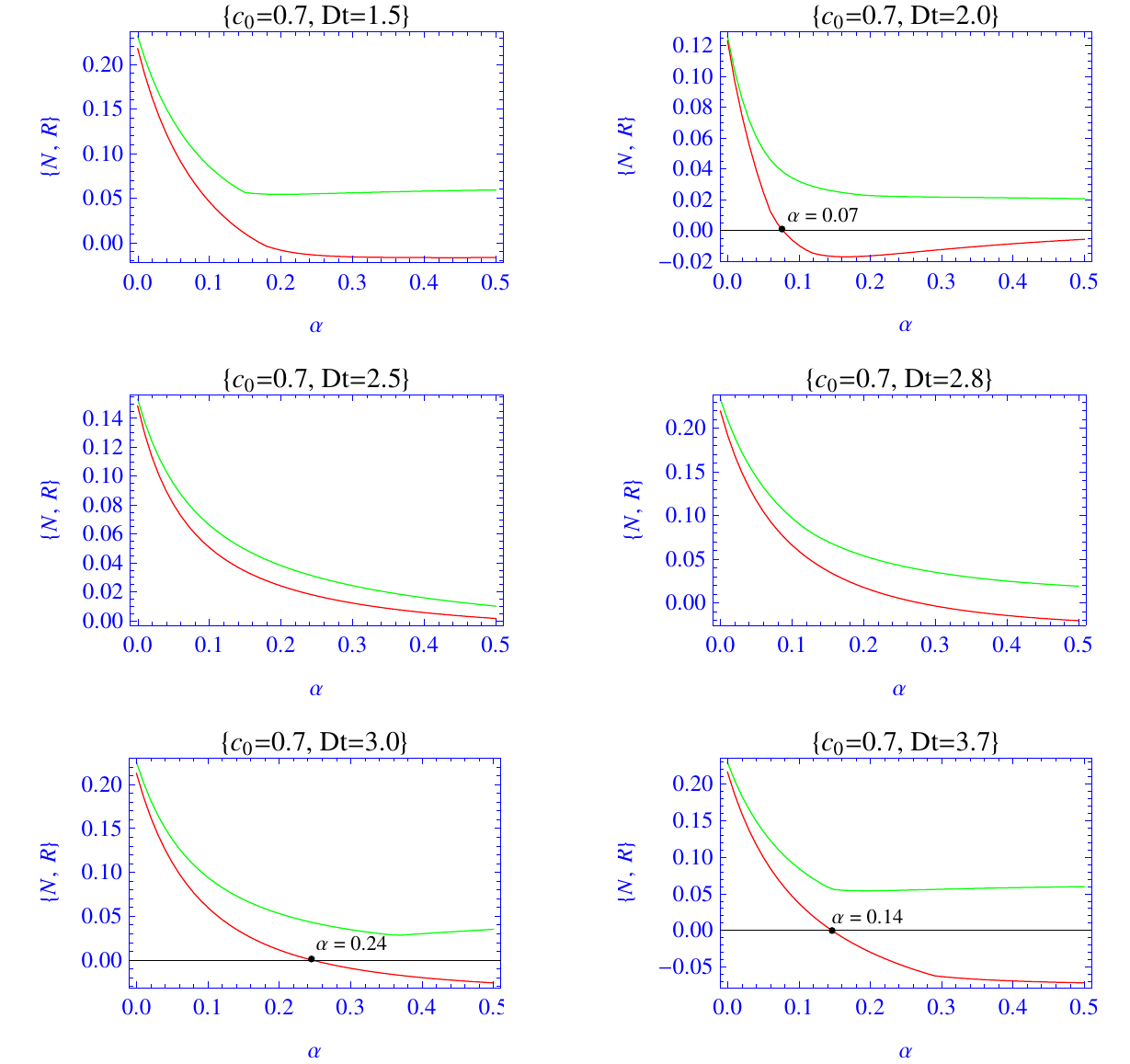}
                \caption{Plot of negativity (N) and realignment criterion (R) with the parameter range $c_{0}=0.7$ and $1.5 \leq Dt \leq 3.7$. Green color graphs represent N and red color graphs represent R.} \label{7}
\end{figure*}
\subsection{Case 2} 
In this case, we consider the state 2 given in subsection 3.2. Here, we again replace $c_{1}^{2}$ in Eq. (\ref{formulae}) in terms of $c_{0}^{2}$ by using the normalization condition given in Eq. (\ref{eq:normal}). Further, we check the distillability of the state by using the reduced density matrix corresponding to this state in the presence of DM interaction. The distillability of the state is detected by using reduction criterion given by Eq. (\ref{r2}). The matrices involved in left hand side of reduction criterion i.e., $(\rho_{A}\otimes I-\rho)$ or $(I\otimes \rho_{B}-\rho)$, incorporate the parameters $Dt$, $c_{0}$ and $\alpha$. The eigenvalues of these matrices have been calculated by taking the range of the parameters as ($3\leq \alpha \leq 4$, $0\leq Dt \leq 5$) with the varying values of parameter $c_{0}$ in the range ($0\leq c_{0} \leq 1$). It is found that the eigenvalues of these matrices achieve negative values and hence the reduction criterion given by Eq. (\ref{r2}) is satisfied for state 2. Our goal is to find out the maximum range of the parameters $Dt$ and $\alpha$ for which the eigenvalues are negative. By simulation, we found these maximum ranges corresponding to the parameter values $c_{0}=0.7$. These eigenvalues are plotted in Fig. \ref{2} with the parameters range $0\leq Dt\leq 5$ and $0\leq \alpha\leq 1.0$ corresponding to the parameter value $c_{0}=0.7$. The corresponding contour plot of the Fig. \ref{2} is depicted in Fig. \ref{3}. 
With the help of this contour plot, we obtained the maximum ranges of the parameters $Dt$ and $\alpha$ for which the state 2 is distillable. These ranges are $1.59\leq Dt\leq 3.75$ and $0\leq \alpha \leq 3.439$. The distillability of the state proves that the states are free entangled. Only free entangled state can be distilled in Bell like pairs, which may be used, further, for quantum information applications. So, based on the  parameters ranges $1.59\leq Dt\leq 3.755$ and $0\leq \alpha \leq 3.439$, we plot the negativity (N) and realignment criterion (R) obtained from Eq. (\ref{formulae}) with different values of the parameter $c_{0}$ in Figs. \ref{5}, \ref{6} and \ref{7}. First, we fix the value of the parameter $c_{0}=0.2$ and vary the value of
$Dt$ within the range $1.5 \leq Dt \leq 3.7$ and for $\alpha$ with $0\leq \alpha \leq 3.439$ and plot the results in Fig. \ref{5}. Observing Fig. \ref{5}, it is concluded that with $\alpha=0$ the realignment criterion achieves the positive values, so initially the state is bound entangled, but as the value of $\alpha$ advances the states become free. So these states can be easily distillable. As the value of the parameter $Dt$ increases, the initial amplitudes of both realignment and negativity fluctuate between $0.04$ and $0.05$. At a particular value of $\alpha=0.02$, the  realignment criterion becomes negative, so it fails to detect the bound entanglement. But corresponding to negative realignment criterion, we can not avoid the possibility of NPT bound entangled states. We observe that for the parameter values, $c_{0}=0.2, Dt=3.0$ the free entanglement in the states vanish after $\alpha=0.12$. 

Next, we increase the value of the parameter $c_{0}$ as $0.4$ and sketch the graphs between $Dt$ and $\alpha$ in Fig. \ref{6}. We found that as the value of parameter $c_{0}$ increases from $0.2$ to $0.4$, the initial amplitudes of both realignment criterion and negativity increases and fluctuates within the limits $0.08$ to $0.10$. Initially the states are bound entangled but as the value of the parameter $Dt$ increases the states become free and hence can be distilled. Again, we found that for some values of $\alpha$, the realignment criterion fails to detect the bound entangled states. These values are $(\alpha=0.07,0.04,0.08,0.06)$. After these values the realignment criterion becomes negative and the possibility of NPT bound entangled states can not be avoided.

Further, we continue our study for $c_{0}=0.7$ and the results are shown in Fig. \ref{7}. Initially the states are bound entangled as the realignment criterion is positive for $\alpha=0$. As the value of parameter $\alpha$ advances, the states become free and hence distillable. The amplitudes of both negativity and
realignment criterion fluctuate between the range $0.12$ to $0.20$. We have found the situation when the states becomes totally free and fully distillable. These parameter values are $(c_{0}=0.7, Dt=1.5)$, $(c_{0}=0.7, Dt=2.5)$ and $(c_{0}=0.7, Dt=2.8)$. However, we detect the values of the parameter $\alpha$ where the realignment criterion becomes negative, these values are $\alpha=0.07, 0.24, 0.14$. After these values of $\alpha$, the realignment criterion fails and corresponding to negative portion of realignment criterion possibility of NPT bound entangled states can not be avoided.
\section{Conclusion}
In this article, we presented a method to distill the bound entangled states by using DM interaction. We consider two qutrits initially prepared in Horodecki bound entangled states and one auxiliary qubit. The auxiliary qubit is prepared in pure state, which interact with any one of the qutrit through DM interaction. By varying the probability amplitude of the auxiliary qubit and DM interaction strength the states can be free and further can be distilled. We have used the reduction criterion, which is necessary condition for distillability, later the maximum values of the parameters of $c_{0}$ and $Dt$ has been obtained for which the the states are distillable. The realignment criterion is used to detect the bound entangled nature of the states and negativity is used to measure the free entanglement in the states. By varying the values of the parameters $c_{0}$, $\alpha$ and $Dt$, the Horodecki bound entangled states can be converted in free entangled states. Free entangled states are easily distillable. We have found that DM interaction can be used to free the Horodecki bound entangled state 2. With certain parameter values of $c_{0}$ and $Dt$, it is completely free and hence distillable. These values are $(c_{0}=0.7, Dt=1.5)$, $(c_{0}=0.7, Dt=2.5)$ and $(c_{0}=0.7, Dt=2.8)$. As the value of the parameter $c_{0}$ increases the initial amplitudes of both negativity and realignment criterion increases. We also found that realignment criterion fails at particular values of $\alpha$ along with the values of $c_{0}$ and $Dt$. So, corresponding to the negative portion of the realignment criterion the possibility of NPT bound entangled states \cite{NPT} exists. We hope that this method of converting bound to free entanglement  can be useful in quantum information processing and varieties of bound entangled states can be checked under DM interaction.
\section*{Acknowledgments}
We would like to thank the anonymous reviewers for going through the manuscript very carefully and suggesting many changes which have greatly enhanced the clarity and presentation of the results.

\end{document}